
\magnification\magstep 1

\def\PRep{Phys.\ Rep.\ }
\def\PR{Phys.\ Rev.\ }
\def\PRL{Phys.\ Rev.\ Lett.\ }
\def\ZP{Z.\ Phys.\ }
\def\NP{Nucl.\ Phys.\ }
\def\PL{Phys.\ Lett.\ }
\def\T{\perp}
\def\pT{p_\T}
\def\qT{q_\T}
\def\hT{h_\T}

\def\Pol{{\cal P}}
\def\vpT{\vec p_\T}
\def\vqT{\vec q_\T}
\def\vhT{\vec h_\T}
\def\vkT{\vec k_\T}
\def\vPol{\vec{\cal P}}

\font\bbf=cmbx10 scaled\magstep2
\pageno=0
\footline{\ifnum\pageno=0 \hfil \else \hss\tenrm\folio\hss\fi}

\hskip 10cm {\bf LYCEN/9423}

\hskip 10cm {\bf TPJU 12/94}

\hskip 10cm {\bf May 1994}
\vskip 2.5cm

\centerline{\bbf Single spin asymmetry in inclusive pion production,}
\vskip 0.2cm
\centerline{\bbf Collins effect and the string model}
\bigskip
\bigskip

\centerline{\bf X. Artru}
\centerline{\it Institut de Physique Nucl\'eaire de Lyon,
IN2P3-CNRS et Universit\'e Claude Bernard,}
\centerline{\it F-69622 Villeurbanne Cedex, France}
\medskip
\centerline{{\bf J. Czy\.zewski}\footnote*{ Presently at \it Institute of
High-Energy Physics, University of Nijmegen, Toer\-nooi\-veld~1,
NL-6525~ED Nijmegen, The Netherlands }}
\centerline{\it Institute of Physics, Jagellonian University,}
\centerline{\it ul. Reymonta 4, PL-30-059 Krak\'ow, Poland}
\medskip
\centerline{\bf H. Yabuki}
\centerline{\it Department of Mathematics, Hyogo University of Teacher
Education,}
\centerline{\it Yashiro, Hyogo, 673-14 Japan}

\vskip 2cm
\centerline{\bf Abstract}
\smallskip

We calculate the single spin asymmetry in the inclusive pion production in
proton-proton collisions.  We generate the asymmetry at the level of
fragmentation function (Collins effect) by the Lund coloured string mechanism.
We compare our results with the Fermilab E704 data from $p\!\!\uparrow\!\!p$
collisions at 200 GeV.  We show that the transversely polarized quark densities
at high Bjorken $x$ strongly differ from these predicted by the SU(6) proton
wave function.
\vfill\eject

\noindent
{\bf 1. Introduction}
\smallskip

\noindent
Quantum Chromodynamics predicts that single transverse spin asymmetries are
suppressed in hard collisions, as a consequence of helicity conservation
(chiral
invariance) in the subprocess.  These asymmetries indeed appear as
interferences
between helicity amplitudes which differ by one unit of helicity, therefore
they
vanish in the limit $m_{\rm quark} \to 0$, or equivalently $Q^2 \to \infty$
($Q$
measures the hardness of the subprocess).  Nevertheless, a number of high $p_T$
reactions persist in showing large asymmetries [1].

These facts do not invalidate QCD but mean that the approach to the asymptotic
regime in $\pT$ is very slow, as regards polarization.  However, in spite of
their ``nonasymptotic'' character, it is not unreasonable to think that the
mechanisms of the asymmetries lie at the parton level.  In other words, the
asymmetries would be manifestations of quark transverse spin (or {\it
transversity}).  Thus, we could extract information from them about the quark
transversity distribution in the nucleon and/or the transversely polarized
quark
fragmentation.  In this paper, we shall present a model for the spin asymmetry
in the reaction
$$
p\!\uparrow + \, p \rightarrow \pi + X
\eqno(1.1)$$
which, unlike previous approaches [2,3], involves the transverse spin asymmetry
of the polarized quark fragmentation [4,5], which hereafter will be refered to
as the Collins effect.

The paper is organized as follows:  section 2 gives a very short review of the
results of Fermilab E-704 experiment.  In section 3, we explain how the Collins
asymmetry can give rise to the observed single spin asymmetry in reaction (1.1)
and deduce lower bounds on the transverse polarizations of the quarks in the
proton, as well as on the size of the Collins effect.  Section 4 presents a
quantitative model based on string fragmentation and section 5 gives the
numerical results.  Section 6 contains discussion of our results and
conclusions.

\bigskip
\medskip
\noindent
{\bf 2. Main features of single spin asymmetry in inclusive pion production}
\smallskip

\noindent
A strong polarization effect has been observed in the reaction (1.1) with 200
GeV transversely polarized projectile protons.  The asymmetry is defined as
$$A_N(x_{F}, \pT) \equiv {\sigma\uparrow - \, \sigma\downarrow
\over \sigma\uparrow + \, \sigma\downarrow},
\eqno(2.1)$$
Assuming that $\uparrow$ refers to the $+ \hat y$ direction (vertical upwards),
and the transverse momentum $\vpT$ of the pion points towards the $+ \hat x$
direction ($\vec p_{\rm beam}$ is along the $\hat z$ axis).  In other words
positive $A_N$ means that for  upward polarization, the pions tend to go to
the left.  The most recent
results, which we will consider in this paper, come from the Fermilab E-704
collaboration.  They were published separately for two kinematical regions:

\item{-}
$x_F > 0$ or fragmentation region.  The asymmetries have been measured for both
charged and neutral pions [6--8].

\item{-}
Central region, $x_F \sim 0$ [9], where the asymmetries were measured for
neutral pions only.  Large asymmetries for high $\pT$ in the central region had
been observed previously also by other experimental groups [10]

\noindent
($x_F$ is the Feynman variable $2p_z^{\rm CM}/\sqrt{s}$ with $p_z^{\rm CM}$
 being the longitudinal momentum of the pion in the CM frame).

In this paper we shall concentrate only on the first case.  In this region the
data show large asymmetry for all pions; positive for $\pi^+$ and $\pi^0$ and
negative for $\pi^-$.  The asymmetries vary from about 0 at $x_F\sim 0.2$ to
about +0.4, +0.15 and $-0.4$, for $\pi^+$, $\pi^-$ and $\pi^-$ respectively, at
$x_F \sim 0.7-0.8$.

\bigskip
\medskip
\noindent
{\bf 3. Hypothesis that E-704 asymmetry is due to Collins effect}

\bigskip
\noindent
{\sl 3.1 Generalities from the parton model.}

\smallskip
\noindent
In the "factorized" parton model (Fig.~1), the cross section for $A+B \to
\pi+X$
in the forward hemisphere is a kind of convolution of a parton distribution
$G_{q/A}(x,\vqT)$, a parton scattering cross section $\hat\sigma_{q+B\to q'+X}
\equiv \hat\sigma_{q\to q'}$ and a parton fragmentation function $D_{\pi/q'}(z,
\vhT)$ .  In short-hand notations,
$$\sigma_{A \rightarrow \pi} \approx G_{q/A}
\otimes \hat \sigma_{q \to q'}
\otimes D_{\pi/q'}
\eqno(3.1)$$
[the factor $ \hat \sigma_{q \to q'}$ may be replaced by $\delta (\vec q- \vec
q\,'\,)$ (no scattering); this is the case of the Dual Parton Model].  Each
factor in this equation may or may not depend on spin.  Transverse polarization
can act at three different levels:

\item{a)} in a dependence of $G_{q/A}(x,\vqT)$ on the azimuth of
$\vqT$ (Sivers effect [2]).

\item{b)} in a single spin asymmetry in $ \hat \sigma_{q \to q'}$
(Szwed effect [3]). In this case, (but not necessarily in case $a$),
the quark $q$ must inherit a part of the polarization of the proton.

\item{c)} in a dependence of $D_{\pi/q'}(z, \vhT)$ on the azimuth of $\vec
h_{\T}$ (Collins effect [4,5]).  Here, a transfer of polarization must occur
not only from the proton to quark $q$ but also from $q$ to $q'$.

\noindent
The first mechanism implies nonzero internal angular momentum inside the
proton,
hence a strong departure from SU(6) model.  As for mechanisms $b$) and $c$),
SU(6) predicts :
$$
A_N^{\pi^-} \simeq -\, {1\over 2}\, A_N^{\pi^+} \qquad
{\rm and} \qquad \left\vert A_N^{\pi^-}\right\vert \le {1 \over 3},
\eqno(3.2)$$
since most of the time, $\pi^{+}$ comes from a valence $u$-quark, $\pi^{-}$
from
a valence $d$-quark of the projectile and the polarizations of $u$ and $d$ are
respectively $+{2\over3}$ and $-{1\over3}$.  These predictions are in conflict
with the E-704 results $A_N^{\pi^+}\simeq -A_N^{\pi^-} \simeq 0.4 $ at large
$x_{F}$.  To conclude, whichever mechanism we choose, SU(6) appears to be badly
violated in $G_{q/p}(x,q_{\T})$ .

\bigskip
\noindent
{\sl 3.2 The Collins effect.}

\smallskip
\noindent
According to Collins [4,5], the fragmentation function of a transversely
polarized quark $q$ takes the form
$$ D_{\pi/q}(\vPol_q,z, \hT) =
\bar D_{\pi/q}(z, \hT)
\left\{ 1 + {\cal A}_{\pi/q} (z, \hT)
\times \vert\vPol_q\vert
\times \sin[\varphi (\vPol_q)-\varphi (\vhT)] \right\}
\,,\eqno(3.3)$$
where $\vPol_q$ is the quark polarization vector ($|\vPol_q| \le 1$), $\vec
h_{\T}$ the pion transverse momentum relative to the quark momentum $\vec q$
and $\varphi (\vec a)$ the azimuth of any vector $\vec a$ about $\vec q$.  The
factors after ${\cal A}$ can be replaced by $|\vec q \times \vhT |^{-1} \
\vPol_q \cdot (\vec q \times \vhT)$.  Such a dependence is allowed by P- and T-
invariance but has not yet been measured.

The Collins effect is the reciprocal of the Sivers effect.  But Collins argued
that the later is prohibited by time reversal invariance [4], while the former
is not.  As for mechanism b), it vanishes for massless quarks due to chiral
symmetry:  single spin asymmetry in $q \rightarrow q^{\prime}$ is not
compatible
with conservation of quark helicity.  Therefore it should be small for hard or
semi-hard scattering.  In conclusion, among the sources of asymmetry a), b) and
c) discussed above, we have a preference for the Collins effect illustrated by
Fig.~1.\footnote{**}{ We shall not discuss other approaches [11] not relying on
the factorized parton description (Eqs.  3.1 or 3.4).  They are not necessarily
in contradiction with the present one.}

\bigskip
\noindent
{\sl 3.3 Consequence for the single spin asymmetry.}

\smallskip
\noindent
Let us consider the hypothesis that E-704 asymmetry is due to the
Collins effect. The polarized inclusive cross section reads
$${d\sigma\over d^3\vec p} = \sum_{\rm flavours\ of\ q,\,k,\,q'} \int dx\, d^2
\vqT\ G_q (x,\vqT) \int dy\, d^2 \vkT\ G_k (y,\vkT)\ \times $$
$$\int d(\cos\hat\theta)\, d\hat\varphi \, {d\hat\sigma^{q+k\to q'+k'}\over
d\hat\Omega} \int dz\, d^2 \vhT\ D_{\pi/q'} (\vPol_{q'},z,\vhT) \, \delta(\vec
p-z\vec q\,'-\vhT) \ ;
\eqno(3.4)$$
the final quark transversity is given by
$$\vPol_{q'} = {\cal R} \ \vPol_{\rm beam} \ {\Delta_{\T}G_q (x,\vqT) \over G_q
(x,\vqT)} \ \hat D_{NN} (\hat\theta) \,.
\eqno(3.5)$$
${\cal R}$ is the rotation about $\vec p_{\rm beam} \times \vec q\,'$ which
brings $\vec p_{\rm beam}$ along $\vec q\,'$,
$$\Delta_{\T} G_q(x,\vqT) \equiv
G_{q\!\uparrow}(x,\vqT) - G_{q\!\downarrow}(x,\vqT)
\eqno(3.6)$$
is the quark {\it transversity distribution} [12,13] in the proton polarized
upwards, and $\hat D_{NN}(\hat\theta)$ is the coefficient of spin transfer
normal to the scattering plane in the subprocess.  Formula (3.1) results from
integration over the target parton variables $y$ and $\vkT$.

At large $x_F$, the dominant quark flavours are $q=q'=u$ for $\pi^+$
production,
$q=q'=d$ for $\pi^-$ production.  Furthermore, the hard scattering occurs
predominantly at small $\hat\theta$ and $\hat D_{NN} (\hat\theta)$ is close to
unity, as in the case of $\hat t$-channel one-gluon exchange [$\, \hat D_{NN}=
-2 \hat s \hat u / (\hat s^2 + \hat u^2) \,$].  Assuming that $\Delta_\T
G_q/G_q$ does not depend on $\vqT$, the results of E704 collaboration imply

$${\Delta_{\T} G_u(\bar x) \over G_u(\bar x) } \times
{\cal A}(\bar z,\bar h_{\T}) \ge {\rm about}\ 0.4
\,,~\eqno(3.7)$$

$${\Delta_{\T} G_d(\bar x) \over G_d(\bar x) } \times
{\cal A}(\bar z,\bar h_{\T}) \le {\rm about}\, -0.4
\,,\eqno(3.8)$$
for $\bar x \bar z \simeq x_F \simeq 0.8$.  $\bar x$ means the most probable
value of $x$.  The inequalities take into account the fact that integration
over
$\vqT$ and $\hat\theta$ allways dilutes the Collins effect.  Thus, we get at
least a lower bound of 0.4 separately for $|\Delta_\T G_u / G_u|$ , $|\Delta_\T
G_d / G_d|$ at large $x$ and $|{\cal A}(z,h_\T)|$ at large $z$ and for the most
probable value of $h_\T$.

\bigskip
\medskip
\noindent
{\bf 4. Simple model of single spin asymmetry}

\smallskip
\noindent

In order to make the conclusions of the previous section more quantitative, we
performed a calculation in a simple model.

We considered only the valence quarks of the projectile proton
with distributions normalized as follows:

$$
\sum_{q=u,d}\int d^2\vqT\int dx\ G_q(x,\vqT ) = 1 \,;
\qquad G_u = 2\,G_d \,.
\eqno(4.1)$$
[$G_q(x)={1\over3}q_{val}(x)$ in conventional notations].  The quark is
accompanied by a diquark carrying the fractional momentum $1-x$ and whose
distribution is
$$
G_{u\!u}(x,\vqT) = G_d(1-x,-\vqT)\,;\ \
G_{u\!d}(x,\vqT) = G_u(1-x,-\vqT)\,.
\eqno(4.2)$$
We did not incorporate any hard or semihard scattering.  Both $\cal R$ and
$\hat
D_{NN}$ of Eq.~(3.5) are equal 1.  The two $q-q\!q$ strings formed after the
collision are parallel to the beam.  We decay them recursively according to the
simple Lund recipe [14].  We use the Standard Lund splitting function:

$$
f(z)=(1+C)(1-z)^C \,,
\eqno(4.3)$$
$z$ being the fraction of the null plane momentum $P^+ \equiv P^0+P^3$ of the
string carried by its leading hadron.  $f(z) = D^{\rm rank=1}(z)$ corresponds
to
the production rate of the first-rank hadron ({\it i.e.}\ the one that contains
the original quark spanning the string).  This gives [14] for all ranks the
inclusive density:

$$
D^{\rm all\ ranks}(z) = (1+C) {1 \over z} (1-z)^C = {1 \over z} f(z).
\eqno(4.4)$$
Thus, for all the other (subleading) hadrons originating from
the string we get:

$$
D^{{\rm rank}\, \ge\, 2}(z) = (1+C) {1-z \over z} (1-z)^C = f(z){1-z \over z}.
\eqno(4.5)$$
The transverse momenta of a quark and an antiquark of a pair created in the
string balance each other (local compensation of the transverse momentum) and
are distributed according to

$$
\rho(\vec \qT)\, d^2\vqT =
{d^2\vqT \over \kappa} \exp \left({-\pi \qT^2 \over \kappa}\right)\,,
\eqno(4.6)$$
$\kappa$ being the string tension.  We took the intrinsic transverse momentum
distribution in $G_q(x,\vqT )$ to be the same as for the tunnelling transverse
momentum in the string:  $G_q(x,\vqT ) = G_q(x)\ \rho(\vqT )$.  It yields an
average intrinsic transverse momentum $\langle\qT^{\rm intrinsic}\rangle =
0.206\,\hbox{GeV}$.

The polarization of the leading quark  $q_0$ is
$$
\vPol_{q_0}(x) =
{ \Delta_\T G_{q_0}(x, \vec {q_0}_\T) \over
G_{q_0}(x, \vec {q_0}_\T)} \cdot \hat y
\eqno(4.7)$$
and is taken to depend only on $x$.

Each quark-antiquark pair created during string breaking is assumed to be in a
$^3P_0$ state (vacuum quantum numbers) [15], {\it i.e.,} with parallel
polarizations.  According to the Lund mechanism for inclusive $\Lambda$
polarization [14], their polarizations are correlated to the transverse
momentum
of the antiquark $\vec {\bar q}_\T$ by

$$
\vPol_q = \vPol_{\bar q} = \,-\,{L \over 1+L} \cdot
{\hat z \times \vec{\bar q}_\T
\over  \vec{\bar q}_\T} \,,
\eqno(4.8)$$
where $L$ is the classical orbital angular momentum of the
$q\bar q$ pair and equals:
$$
L = { 2\ \bar q_\T \sqrt{m_q^2 + \bar q_\T^2} \over \kappa }
  \simeq { 2\ \bar q_\T^2 \over \kappa },
\eqno(4.9)$$
(see Fig.~2 for a schematic explanation).

In order that $q_0$ and $\bar q_1$ of Fig.~2 combine into a pion, they have to
form a spin singlet state, the probability of which is
$$
{1 \over 4} \, (1 - \vPol_{q_0} \cdot \vPol_{\bar q_1}) \,,
\eqno(4.10)$$
in accordance with the projector on the singlet state
${1 \over 4} - \vec s(q_0) \cdot  \vec s(\bar q_1)$.
The factor (4.10) causes the Collins effect:  if $q_0$ in Fig.~2 is polarized
upwards then $\bar q_1$ and the pion which contains $\bar q_1$ tends to go to
the left-hand-side of the $\hat z$ direction.

Vector mesons are ignored (accordingly, we omit the factor ${1\over4}$ in
Eq.~(4.10)).  In our model the asymmetry for vector mesons would be of the
opposite sign but three times smaller in magnitude when compared to that of the
scalar ones.  The observed asymmetry of the pions resulting from decays of the
vector mesons would be even smaller, due to integration over decay
angles\footnote{$^{\dag}$}{
It has been shown however that the vector meson can also have
a {\it tensor} polarization [16]
which would result in the Collins effect for the decay products. We did not
include this possibility. Another source of asymmetry [5]
could be the interference between the vector meson
and the nonresonating background.}.
Anyway, at large $x_F$, there are mostly direct pions.

We do not introduce the Collins effect in subleading ranks or in the
fragmentation of the diquark.  For the fragmentation function (4.3), the
probability that the detected pion is the leading one is equal to its momentum
fraction $z$.  Hence, the model can be a good approximation mostly for pions of
high positive $x_F$.  We provide a short discussion of this point in the last
section.  We shall divide the sources of the pion production into three parts:

\item{1. }
the pion is the leading particle of the string spanned by the quark $q_0$.  The
Collins effect (and the asymmetry) appears only in this contribution.

\item{2. }
The pion is a subleading particle of the string spanned by that quark, or

\item{3. }
it is a subleading particle of the string spanned by the diquark accompanying
the quark $q_0$.  The leading particle of the string in this case is a baryon.
Taking into account that $x_F \approx xz$ and that the transverse momentum of
the pion is the sum of the transverse momenta of its constituents we get the
production rates for all the three cases:

$${1 \over \sigma_{\rm tot}} \left[ {d\sigma \over dx_F d^2\vpT} \right]_{\rm
quark}^{\rm rank=1} = \sum_{q=u,d}\int dx\, dz\, d^2 \vqT\, d^2 \vec{\bar q}_\T
G_q(x, \vqT)\ c_1\ D^{\rm rank=1}(z)$$

$$\times \left(1 - \vPol_q \cdot \vPol_{\bar q}(\vec{\bar q}_\T)\right)\
\rho(\vec{\bar q}_\T)\ \delta(x_F - xz)\ \delta^2(\vpT - \vqT - \vec{\bar
q}_\T)
\eqno(4.11)$$

$$ {1 \over \sigma_{\rm tot}} \left[ {d\sigma \over dx_F d^2\vpT} \right]_{\rm
quark}^{{\rm rank}\, \ge\, 2} = \sum_{q=u,d}\int dx\, dz\ G_q(x)\ c_2\ D^{{\rm
rank}\, \ge\, 2}(z)\ \rho_\pi(\vpT)\ \delta(x_F - xz)
\eqno(4.12)$$

$$ {1 \over \sigma_{\rm tot}} \left[ {d\sigma \over dx_F d^2\vpT} \right]_{\rm
diquark}^{{\rm rank}\, \ge\, 2} = \sum_{q=u,d}\int dx\, dz\ G_{q\!q}(x)\ c_3\
D^{{\rm rank}\, \ge\, 2}(z)\ \rho_\pi(\vpT)\ \delta(x_F - xz)
\eqno(4.13)$$
and the total production rate is the sum of the three:

$$ {d\sigma \over dx_F d^2\vpT} = \left[ {d\sigma \over dx_F d^2\vpT}
\right]_{\rm quark}^{\rm rank=1} + \left[ {d\sigma \over dx_F d^2\vpT}
\right]_{\rm quark}^{ {\rm rank}\, \ge\, 2} + \left[ {d\sigma \over dx_F
d^2\vpT} \right]_{\rm diquark}^{ {\rm rank}\, \ge\, 2}
\eqno(4.14)$$
$\rho_\pi(\vpT)$ is the convolution of $\rho(\vqT)$ and $\rho(\vec{\bar q}_\T)$
and is also Gaussian but of twice larger variance\footnote{$^{\ddag}$}{
We did not follow exactly the $\delta(\vec p-z\vec q\,'-\vhT)$ prescription
of (3.4) ; in other words we gave all the transverse momentum of
the leading quark to the first rank particle. The resulting error is small at
large $x_F$.}.
$c_1$, $c_2$ and $c_3$ are flavour factors and correspond to the probability
that $q$ and $\bar q$ match to form a pion of the appropriate charge.

Since the production rate varies with the azimuthal angle of the pion momentum
$\phi$ like in (3.3) then, in order to obtain the asymmetry at given
values of $\pT$ and $x_F$, we need to compare $d\sigma (x_F,\pT,\phi)$ only at
$\phi = 0$ and $\phi = \pi$:
$$
A_N(x_F,\pT) = {d\sigma (x_F,\pT,0) - d\sigma(x_F,\pT,\pi) \over
                d\sigma(x_F,\pT,0) + d\sigma(x_F,\pT,\pi)}
\eqno(4.15)$$
which completes the calculation.

\bigskip
\medskip
\noindent
{\bf 5. Numerical results}

\smallskip
\noindent
For the numerical calculations we parametrized the quark distributions as
 follows:
$$
G_u(x) = 2\ G_d(x) = {5\over 2}\ x^{1/2}\ (1-x).
\eqno(5.1)$$
We have used the string tension $\kappa = 0.17$ GeV$^2$ and the parameter of
the
fragmentation function $C = 0.3$.  In pair creation we have used the flavour
abundances with the ratio $u :  d :  s = 3 :  3 :  1$, which determine the
coefficients in Eqs (4.11--4.13)  to be $c_1 = 3/7$, $c_2 = c_3 = 9/49$ for
charged and $c_2 = c_3 = 18/49$ for neutral
pions\footnote{$^{\S}$}{
In this model, every $u\bar u$ or $d\bar d$ meson is considered as a $\pi^0$
(no
$\eta^0$) ; it gives $\sigma(\pi^+)+\sigma(\pi^-) = \sigma(\pi^0)$, instead of
$2\ \sigma(\pi^0)$ as required by isospin.  Nevertheless, Eq.  (5.3) below
remains true.}.

In Fig.~3a we show our results compared to the data [7] of $0.7 < \pT <
2.0\,\hbox{GeV}$.  The dotted, dashed and full lines correspond to various
dependences of the quark polarization $\Pol_q$ on the momentum fraction $x$:
constant, proportional to $x$ and to $x^2$ respectively.  Results obtained for
all the three choices converge at $x_F = 1$.  The maximal values of the
polarization, reached at $x=1$, are those motivated by the SU(6) wave function
of the proton:  $\Pol_u = +2/3$ and $\Pol_d = -1/3$.

One sees that the resulting asymmetry $A_N$ strongly disagrees with the data
for
the negative pions.  This confirms the conclusion of the Section 2.  The
measured asymmetries in $\pi^-$ production are, for $x_F > \sim 0.5$, equal in
the absolute value but of the opposite sign to those of $\pi^+$.  This cannot
be
accounted for by SU(6) where the asymmetry of $\pi^-$ is negative but twice
smaller than that of $\pi^+$.

The results compared to the data at lower transverse momenta, $0.2 < \pT <
0.7\,\hbox{GeV}$, are shown in Fig.~3b.  Here no disagreement with the data is
seen.  The data do not reach however as high values of $x_F$ as in the high
$\pT$ interval.

The behaviour of the measured asymmetries at high $x_F$ and $\pT$ encouraged us
to try out a parametrization with maximal possible transverse polarizations of
quarks one could choose in the proton:  $\Pol_u =1$ and $\Pol_d = -1$ at $x=1$.
The results are plotted in Figs.  4a and 4b for both $\pT$ intervals.  The
constant polarizations $\Pol_q(x) = \hbox{const}$ (dotted lines) result in too
large asymmetry, at least at small and intermediate $x_F$.  This suggests that
the transversely polarized quark densities fall down at small Bjorken $x$
values.

The full lines follow the data quite well.  Only the measured asymmetries of
$\pi^-$ slightly differ from our results at small values of both $x_F$ and
$\pT$.  One might conclude that $\Delta_\T G_q(x)/ G_q(x)$ varies with $x$
somwhere around $x^2$ but such conclusion is dangerous in scope of the
simplicity of the model.

In order to check how our results depend on the shape of the quark distribution
$G_q(x)$ we did the calculation also for
$$
G_q(x) \sim x^{-1/2},
\eqno(5.2)$$
which differs strongly from (5.1) but gives better account of the leading
baryonic effect ($G_{q\!q}(x)\sim (1-x)^{-1/2}$ and the diquark tends to carry
a
substantial momentum fraction of the proton).  The comparison is shown in
Fig.~5.  One sees that for the distribution (5.2) the asymmetry (dashed line)
is
slightly smaller at large $x_F$ but the difference is not large.  The full line
comes from Fig.~4 and corresponds to the distribution (5.1).

In the parton model the $\pi^0$ asymmetry is just a combination
of the $\pi ^+$ and $\pi^-$ ones :

$$ A_N(\pi^0) = {\sigma(\pi^+)\ A_N(\pi^+) + \sigma(\pi^-)\ A_N(\pi^-)
\over \sigma(\pi^+) + \sigma(\pi^-) } \,.
\eqno(5.3)$$

Nevertheless, we show in Fig.~6 a comparison of our results to the E704 data
[6]
on $\pi^0$ production in $pp$ and $\bar pp$ collisions.  The curve obtained
with $\Pol_q(x)\sim x^2$ (full line) agrees with the data also here.  The lower
$\pT$ bound in the $\pi^0$ case varies from 0.5 to 0.8 GeV depending on $x_F$
and was taken into account in our calculation.  This is the reason of a bit
wiggly shape of the lines.  For the neutral pions the difference between SU(6)
and the maximal polarizations of the quarks in the proton cancels and there is
no difference in our predictions there.

Finally, in Fig.~7 we plot the $\pT$ dependence of the asymmetry of $\pi^0$.
The agreement with the data is also good.  The shape of the curves reflects the
Lund parametrization (4.8).  Only the rise of the asymmetry at high $\pT$ and
close to the central region of $x_F$ cannot be described by the model.  We
believe that this rise can result only from a combination of hard scattering
with the Collins effect.  The former has not yet been included in our
calculation.

\bigskip
\medskip
\noindent
{\bf 6. Discussion and conclusions}

\smallskip
\noindent
To summarize, we calculated the single transverse spin asymmetry in high-energy
$pp$ collisions in a simple model involving the Collins effect (asymmetry
arising at the level of fragmentation of a quark into hadrons).  We
parametrized
the Collins effect by the Lund mechanism of polarization in the coloured string
model.

We got good agreement with the data when we assummed that:

\item{a)}
The transverse polarization of the $u$ and $d$ quarks in the transversely
polarized proton are close to unity but of the opposite sign ($\vPol_u =
\vPol_{\rm proton}$, $\vPol_d = - \vPol_{\rm proton}$) at momentum fraction $x$
close to 1.

\item{b)}
The dependence of the quark transversity (or polarization) on the momentum
fraction $x$ is close to be proportional to $x^2$.

The conclusion a) is model-independent provided that the asymmetry arises in
the
quark fragmentation (Collins effect), which is a reasonable assumption as
argued
in section~2.

The quark transversities we got :
$$
{\Delta_\T G_u(x)\over G_u(x)}\approx
\,-\,{\Delta_\T G_d(x)\over G_d(x)} \approx 1 \qquad ({\rm for}\ x\to 1)
\eqno(6.1)$$
are, in fact, not unreasonably large. Consider a covariant model of the
baryon consisting of a quark and a bound spectator diquark [13,17] ;
then
$$
G_{q \uparrow /B \uparrow}(x)= {x \over 16\pi^{2}}
\int_{-\infty}^{q^2_m} dq^2  \left({g(q^2) \over
q^2-m^2_q}\right)^2
\sum_{\rm diquark\ polarization}
\left|\bar u(q \uparrow) V u(p \uparrow)\right|^2
\eqno(6.2)$$
where $g(q^2)$ is the $q-q\!q-B$ form factor, $V=1$ for a scalar diquark,
$V=\gamma_5 \gamma \cdot \varepsilon$\  for a $1^+$ diquark of polarization
$\varepsilon ^\mu$ and
$$
q^2_m= xm^2_B-{x \over 1-x} m^2_{q\!q} \,.
\eqno(6.3)$$
Formula (6.2) is similar to the covariant Weizs\"acker--Williams formula, but
for a ``spin ${1 \over 2}$ cloud'').  Independently of $g(q^2)$, this model
predicts the following behaviours at $x\to 1$:

\item{-}
for a $1^+$ spectator diquark, helicity is fully transmitted [~$\Delta_L G_q(x)
/G_q(x)\to 1$~], transversity is fully reversed [~$\Delta_\T G_q(x) / G_q(x)
\to -1$~].  In particular, $\Pol_d(x) \to -1$.

\item{-}
for a $0^+$ diquark, $\Delta_\T G_q(x)$ and $G_{q^+}(x)$ coincide and, for
$g(q^2)$ decreasing faster than $q^{-2}$, they exceed ${2\over3} G_q(x)$ as
$x\to 1$

\noindent
Thus, a dominance of the scalar spectator for the $u$ quark and the
pseudo-vector one for the $d$ at $x\sim 1$ could lead to the large opposite
transversities as in (6.1).

The conclusion b), related to the $x_F$ dependence, is model-dependent and
cannot be taken too seriously.  One needs a good parametrization of the Collins
effect before one can deduce the $x$ dependence of the quark transversity.  Our
parametrization is an approximation which should work only at reasonably high
values of both $x_F$ and $p_T$.  We took into account the Collins effect only
for the first-rank (leading) hadrons, wherefrom ${\cal A} \propto z$ in (3.3).
The second-rank hadrons have the asymmetry of the opposite sign as compared to
the first-rank ones.  More generally, the subsequent ranks are asymmetric in
the
opposite way to each other (as required also by local compensation of
transverse
momentum).  This should cause a faster decrease of ${\cal A}$ at lower $z$
values, where the higher-rank hadrons are more important.  Unfortunatly this
feature was not possible to include in our simple semi-analytical calculation,
since the yields of rank-2 (and higher) hadrons do not have simple analytical
forms.  The contribution of vector mesons also should reduce ${\cal A}$ at
lower
$z$.  Assuming $\bar x\sim \bar z\sim \sqrt{x_F}$, a steeper $ {\cal A}$ (for
instance $\propto z^2$) would imply a flatter $\Pol_q(x)$ (for instance
$\propto
x$).

In our model the high $\pT$ hadrons originate from the tail of the Gaussian
distribution of the intrinsic and tunneling transverse momenta.  Incorporating
the hard scattering should not change the results very much at large $\pT$.
There will be some smearing of $\vhT$, hence a reduction of the asymmetry, but
not too severe because of the "trigger bias" effect:  in most high $\pT$
events, the contributions of string decay and of hard scattering to the
transverse momentum are rather large and point in approximately the same
direction, as in Fig.  1 (in order to sum up to the large $\vpT$ of the pion).
At small $\pT$, the trigger bias effect would work less and this could improve
the agreement with the data in that region (Fig.~4b).

One main conclusion of this paper is that single spin asymmetry may be the
first
experimental indication for the existence of the Collins effect.  A more
detailed experiment would be usefull to select between this and alternative
explanations.  Besides its theoretical interest, the Collins effect may be the
most efficient "quark polarimeter" necessary for the measurements of the
transversity distributions in the nucleons [5,18]. We hope that this effect
will
soon be tested directly, for instance in the azimutal correlation of two pion
pairs from opposite quark jets in $e^+ e^-$ annihilation.

\bigskip
\medskip
\noindent
{\bf Acknowledgements}
\smallskip

\noindent
We are grateful to J.~Szwed for discussions. X.A. and J.C. acknowledge the
financial support from the IN2P3--Poland scientific exchange programme.
J.C. has been also supported by the Polish Government grants of KBN no.\
2~0054~91~01, 2~0092~91~01 and 2~2376~91~02 during completion of this
work.

\bigskip
\medskip
\noindent
{\bf References}
\smallskip

\item{[1]} K.~Heller, $7^{\rm th}$ Int.\ Conf.\ on Polarization Phenomena in
Nuclear Physics, Paris 1990, p.\ 163, and references therein; P.R.~Cameron {\it
et al.}, \PR {\bf D32}, 3070 (1985); T.A.~Armstrong {\it et al.}, \NP
{\bf B262}, 356 (1985); S.~Gourlay {\it et al.}, \PRL {\bf 56}, 2244 (1986);
M.~Guanziroli {\it et al.}, \ZP {\bf C37}, 545 (1988)

\item{[2]} D.~Sivers, \PR {\bf D41}, 83 (1990); \PR {\bf D43}, 261 (1991)

\item{[3]} J.~Szwed, Proc.\ of the 9$^{\rm th}$ International Symposium ``High
Energy Spin Physics'' held at Bonn, 6--15 Sep.\ 1990, Springer Verlag 1991;
\PL {\bf B105}, 403 (1981)

\item{[4]} J.~Collins, \NP {\bf B396}, 161 (1993)

\item{[5]} J.~Collins, S.F.~Heppelmann and G.A.~Ladinsky, PSU/TH/101, April
1993

\item{[6]} D.L.~Adams {\it et al.}, \PL {\bf B261}, 201 (1991)

\item{[7]} D.L.~Adams {\it et al.}, \PL {\bf B264}, 462 (1991)

\item{[8]} D.L.~Adams {\it et al.}, \ZP {\bf C56}, 181 (1992)

\item{[9]} D.L.~Adams {\it et al.}, \PL {\bf B276}, 531 (1992)

\item{[10]} J.~Antille {\it et al.}, \PL {\bf B94}, 523 (1980); V.D.~Apokin
{\it
et al.}, \PL {\bf B243}, 461 (1990); B.E.~Bonner {\it et al.}, \PR {D41}, 13
(1990); S.~Saroff {\it et al.}, \PRL {\bf 64}, 995 (1990)

\item{[11]} B.E.~Bonner {\it et al.}, Phys.\ Rev.\ Lett.\ {\bf61}, 1918 (1988);
M.G.~Ryskin, Sov.\ J.\ Nucl.\ Phys.\ {\bf48}, 708 (1988); M.S.~Amaglobeli {\it
et al.}, Sov.\ J.\ Nucl.\ Phys.\ {\bf50}, 432 (1989); C.~Boros, Liang Zuo-tang
and Meng Ta-chung, \PRL {\bf70}, 1751 (1993); H.~Fritzsch, Mod.\ Phys.\ Lett.\
{\bf A5}, 625 (1990)

\item{[12]} J.P.~Ralston and D.E.~Soper, \NP {\bf B152}, 109 (1979);
J.L.~Cortes, B.~Pire and J.P.~Ralston, \ZP {\bf C55}, 409 (1992);
R.L.~Jaffe and Xiangdong Ji, \NP {\bf B375}, 527 (1992)

\item{[13]} X.~Artru and M.~Mekhfi, \ZP {\bf C45}, 669 (1990)

\item{[14]} B.~Andersson, G.~Gustafson, G.~Ingelman, T.~Sj\"ostrand, \PRep\
{\bf
97} 31 (1983)

\item{[15]} A.~Le~Yaouanc, L.~Oliver, O.~P\`ene and J.-C.~Raynal, {\it Hadron
Transitions in the Quark Model} (Gordon and Breach, 1988)

\item{[16]} Xiangdong Ji, \PR {\bf D49}, 114 (1994)

\item{[17]} H.~Meyer and P.J.~Mulders, \NP {\bf A528}, 589 (1991)

\item{[18]} X.~Artru, QCD and High Energy Hadronic Interactions
(Ed.~J.~Tr\^an Thanh V\^an, Editions Fronti\`eres, 1993), p.~47

\bigskip
\medskip
\noindent
{\bf Figure captions}

\smallskip

\item{Fig.~1} Inclusive pion production.  Two events (a) and (b), symmetric
with
respect to the $yz$ plane, are represented.  Without polarization, they would
have the same probability.  In the polarized case, the Collins effect favours
the case (a).

\item{Fig.~2} Production of the leading pion in a string drawn by a
transversely
spinning quark.

\item{Fig.~3} Single spin asymmetry measured by E704 collaboration.  The curves
are our model results calculated with quark transverse polarizations at $x = 1$
as in SU(6) wave function of the proton ($\Pol_u = 2/3$ and $\Pol_d = -1/3$).
The data are from Ref.~[7].

\item{Fig.~4} The asymmetry calculated under an assumption that the quark
polarizations at $x = 1$ are:  $\Pol_u = +1$ and $\Pol_d = -1$.  The data are
as
in Fig.~3.

\item{Fig.~5} Comparison of two various quark distributions.  The full lines
and
the data are as in Fig.~4.

\item{Fig.~6} The asymmetry of $\pi^0$'s.  Here there is no difference between
the predictions of SU(6) and the maximal polarizations.  The quark
distributions
are as in Figs~3 and 4.  Data come from Ref.~[6].

\item{Fig.~7} $\pT$ dependence of the $\pi^0$ asymmetry in two intervals of the
Feynman variable $x_F$. Data are from Ref.~[8].

\end